# Applying a Differential Evolutionary Algorithm to a Constraint-based System to support Separation of OTDR Superimposed Signal after Passive Optical Network Splitters

Lima, Gerson F. M; Lamounier, Edgard; Barcelos, Sergio; Cardoso, Alexandre; Peretta, Igor; Muramoto, Willian; Barbara, Flavio

*Abstract*— The FTTH (Fiber To The Home) market currently needs new network maintenance technologies that can, economically and effectively, cope with massive fiber plants. However, operating these networks requires adequate means for an effective monitoring cost. Especially for troubleshooting faults that are associated with the possibility of remote identification of fiber breaks, which may exist in the network. This is of great value for operators. Optical Time Domain Reflectometry (OTDR) techniques are widely used in point-to-point optical network topologies. Nevertheless, it has major limitations in tree-structured PONs (Passive Optical Networks), where all different branches backscatter the light in just one conventional OTDR trace with combined signals arriving on the OLT (Optical Line Terminal) side. Furthermore, passive power splitters used in FTTH networks input large attenuation, impoverishing the reflected signal. This makes the identification of the very branch affected by the problem practically impossible, when considering conventional analyses [32]. From this scenario, arrives the PON network operator's requirement for an innovative solution that allows the precise identification of the affected branch. The use of constraint-based techniques have been applied in a large amount of applications for Engineering Design, where the duties imposed for graphics and equations constraints result in valued features to CAD/CAE software capabilities. Currently, it provides a faster decision making capacity for engineers. This work applies the constraint-based approach along with a Differential Evolutionary Algorithm to separate the superimposed OTDR signals, after the splitters of a FTTH Passive Optical Networks. This research introduces a new set of algorithms performing a coupling to an Optical Network (ON) CAD Design with its correspondent OTDR measurement signal, considering its geographical distribution branches of different lengths after the splitter. Results of this work are presented in a FTTN (Fiber To The Node) prototype arrangement, using a 1:8 passive power splitter.

**1-INTRODUCTION**

Over the last 30 years, optical fiber communication technology has come down in price. Essentially, long distances and the metropolitan telecommunication market have fostered it. Thus, by considering economic feasibility, technical superiority and market demand, optical fibers will definitively replace the twisted pairs in home access networks. Moreover, Fiber to the Home (FTTH) technology has arrived. Passive Optical Networking (PON) is an economically viable technology for providing ultra-broadband FTTH access networks [1]. This viability is achievable through the shared network infrastructure model employed. A non-shared infrastructure model, in which a dedicated fiber runs from the central office to each home, multiplies the engineering costs, instead of sharing it. It also demands terminals that are more active, more space in the central offices and more energy supply, which therefore increases risk of failure.

Therefore, the PON model has become the choice for providing Fiber to the Home (FTTH) networks. A FTTH plant will gradually replace the copper wire pair network. Optical fibers will then be in the service of an enormous number of homes. This poses a great level of concern: how to efficiently and economically speed up the network test? In addition, how to repair the massive FTTH network plants in the future? A variety of Optical Time Domain Reflectometry (OTDR) testing methods have been designed for the verification and troubleshooting of PON's. These include testing all points from the optical network terminal (ONT) to the central office (CO). Furthermore, other methods include testing the feeder part of the network from the CO, and in some cases, simply not testing at all [2].

In fact, OTDR methods provide reliable results. Besides, since it is a single-ended method, it significantly reduces staff time. One perceived drawback is the cost of the OTDR instrument when added to the required user skill level. In addition, reflectometric signals are processed by software using different methodologies to recognize events, failures and degradations in the optical fiber. Point-to-point fiber networks [3] have dominated this matter for some time. However, for FTTH PON architectures, several new challenges arise when testing using optical passive power splitters [4].

One solves this problem by either testing from the ONT side or testing up to the splitter position only.

The former is a non-economical solution, as it would require an enormous amount of testing instruments and well-trained field technicians [5]. The latter can only see the feeder part of the network, i.e., from the CO up to the splitter position, which is usually less prone to network`s failure portion. After the splitter, the OTDR measurement brings little information because the Rayleigh scattering is strongly attenuated by the double-passage through the splitter [6]. This is because the 64 backscattered signals, from the PON branches, superimpose the OTDR receiver. This makes it impossible to recognize events from single PON branches. Besides, traditional OTDR technology has a low dynamic range when used with short OTDR pulses, which is a requirement for PON measurements.

PON using tree topology or P2MP configuration is one of the most promising solutions for fiber-to-the-office (FTTO), fiber-to-the-business (FTTB), fiber-to-the-curb (FTTC), fiber-to-the node (FTTN) and FTTH, since it breaks through the economic barrier of traditional P2P solutions. The P2MP cable plant provides branching optical paths from the telecommunications operators switching equipment (Optical Line Terminal – OLT) to more than one contiguous location. Thus, portions of the optical paths are shared by traffic to and from multiple locations [6].

Passive optical splitter (power splitting element/branching device) or optical coupler is a device used to broadcast an optical signal from one fiber to many others. In the most general case, the optical splitter is configured as 1xN – 1 input port to N output ones. Optical signals on the input port are branched to all output ports. An optical splitter may also be used to multiplex the optical signals from several fiber lines onto a single line. However, using optical coupler poses several problems in the OTDR testing [7]. If the OTDR is connected to the N side of a 1xN optical splitter, then the waveform shows a large loss, limiting OTDR's ability to test far beyond the optical splitter. Furthermore, when connected to the input side of a 1xN optical splitter, the waveform shows a smaller drop than when the OTDR is connected to one of the N output ports. Then, a higher signal level results due to the combined signatures of all N fibers on the branching side of the optical coupler, which are superimposed and makes it very difficult (sometimes impossible) to associate events on the waveform with the specific lines on which events occur. This is probably the most common problem faced by technicians when dealing with an optical passive power splitter. Some solutions arise, such as the fiber fault, monitoring technique, as proposed by C.H. Yeh [8] and fiber-break monitoring system, as proposed by A.A.A. Bakar [9].

In this work, we have developed a system to integrate the graphical network design (CAD-Computer Aided Design / GIS-Geographical Information System) with the OTDR measures made from CO, using an in service dispositive. In the same solution set, the authors implemented a Differential Evolutionary Algorithm [1, 15] to obtain a signal separation for each branch of the network, associating events on the OTDR signal waveform with the specific cable design lines in which the events occur on a map. This work used an arrangement created, specifically, to investigate a concept proof to determine an OTDR Splitter relation between separated branches. To do so, we propose a superimposed mathematics equation, which promotes the simulation of OTDR signal after the PON splitter.

The motivation behind this work is to separate the OTDR signal in order to plot, over a network design map (CAD), the separated signal and use it, to provide location based on information to improve maintenance procedures. Therefore, the signal separation is very important for providing an innovative tool that joins the OTDR fiber signal separated after passing the splitter coupled with GIS design information. The proposed methodology is shown in Figure 1.

From the FTTN PON Design (CAD/GIS), the length of each fiber cable that goes on to constitute the optical network, after the splitter, is obtained, providing the expected distance where Fresnel´s reflection [6] occurs. From the physical network, the OTDR provides the acquired signal record in a database and the Differential Evolutionary Algorithm simulates each channel after the splitter. In this work, the authors propose carrying out these measurements during the network installation phase using CAD Design parameters. Finally, the separated signal for each channel is coupled to the Network Design in a Map (Figure 1). In the following sections, each step is described in details.

**2-DERIVATION OF THE SPLITTER EQUATION**

**2.1-Experimental Setup**

The authors designed the tested network, presented in Figure 2, to serve as a platform for studding the mechanisms and characteristics of an optical signal in working (good/ideal) condition, similar to a Fiber-To-The-Node arrangement (FTTN). This setup mounts a laboratory OTDR in a location as if it were inside a CO (Central Office), distributing signals to multiple ONUs (Optical Network Units) at different locations of residential customers (in the downstream direction), bypassing the 1x8 optical splitter. Table 1 presents further network details.

Table 1 - Bobbin details connected in the branches.

| Branch | Insertion Loss (dB) | Optical Return Loss (dB) | Length (Km) | Loss/Km (dB/Km) |
|---|---|---|---|---|
| 0 | - | - | 2.5420 | 0.2 |
| 1 | 10.375 | 71.14 | 5.6578 | 0.190 |
| 2 | 12.173 | 59.61 | 6.0108 | 0.224 |
| 3 | 11.438 | 60.14 | 5.7039 | 0.200 |
| 4 | 10.088 | 73.25 | 3.3251 | 0.233 |
| 5 | 10.804 | 72.18 | 2.5168 | 0.193 |
| 6 | 9.684 | 78.01 | 2.6294 | 0.219 |
| 7 | 11.219 | 71.94 | 19.4239 | 0.203 |
| 8 | 10.50 | 72.65 | 12.4718 | 0.223 |

The standard ITU recommends a 1625nm laser source for in-service monitoring. However, in this stage, we used a commercially available OTDR model (MW9076D1– Anritsu) with a 1550nm laser source. Therefore, for concept proof, this specification should not make any relevant difference, because in this experiment we used single-mode optical fiber [24] with a low bending sensitivity class. When subjected to one complete turn, a mandrel with a diameter of 20 mm should not exhibit attenuation variation exceeding 0.75 dB at 1550 nm and 1.5 dB at 1625 nm. On the other hand, since the lengths of the tested network branches varies from 5 to 12km, the measurements in 1550nm and in 1625nm are very similar to the initial proof of concept experiments. The fiber has attenuation slightly higher in 1625 than in 1550nm (0.3dB/km). As a result, the slope will be slightly higher. Another influence of the wavelength is the refraction index. In our experiments, the events occur slightly dislocated (few meters).

The authors chose this configuration as it represents a common architecture. In this layout, a primary 8-port splitter is installed, leaving secondary fibers going to the secondary network until the last mile, where one can get the OTB (Optic Termination Boxes) / NAP (Network Access Point). Each fiber branch at its end is cleaved, which provides 11dB of reflectance.

**2.2 Measurement parameters**

Few parameters need for defining the OTDR setup in order to get best results [26]. Distance range (DR) also known as display range. It limits the amount of displayed fiber. The distance range affects the test accuracy and the time required to complete tested fiber. Measurement resolution is the space between data points. In our experiment, we defined it with 0.5 meters. Pulse Width is the duration of the laser pulse. By selecting a longer or shorter pulse width, we can control the amount of backscatter level coming back along with the dead zone size. In this case study, we used Pulse Width of 500ns, about 100m of resolution, allowing a good measurement for the total length of the proposed experimental arrangement of this work. However, it is too inaccurate for multiple dwelling unit's case due the small distances involved.

In future experiments, we plan to use a pulse width of 10ns or less to allow the separation of elements 2m apart. Nevertheless, with this pulse width it is not possible to evaluate the signal after the splitter, only the reflective events of the cleaved fibers immerse into the noise.

The amount of averages is a key concept to improve the signal-to-noise ratio (SNR) of the trace and to get a more a reliable and smoother-looking trace. In fact,

OTDR send out thousands of measurement pulses every second. Every pulse provides a set of data points, and then their average is joined with subsequent sets of points. Usually, a lot of averaging is required and takes time. We have defined the amount of averaging needed to provide consistent results.

The conjunction of data to allow the evaluation of results and simulations is presented in the sequence of tests forward. We configured the OTDR using the following parameters:

- DR (Distance Range): 25Km (it should have the smallest value greater than the maximum length of the network, in this case 25km);
- Pw (Pulse Witdh): 500ns (smallest pulse width which allowed viewing the fiber end);
- Avg (Averages): 60 (smaller number of averages needed to observe the fiber signal with minimal noise, researcher´s discretion);
- The OTDR generates traces for each recorded measurement into a CSV (Comma Separated Value) text file (measurement resolution 0.5 meters).
- 

### 2.3- Test sequence

The undertaken test sequence has the purpose of providing references to the algorithm development and concept proof. It does not mean that these tests should be performed at the time of the network installation, since such measures would be unfeasible in practice. We are proposing this sequence of tests to prove the development affirmed in this study.

**Sequence A** is generated by analyzing each separate channel of the optic fibers after the splitter, thus unplugging the others (Table 2a):

Table 2a – Tests with each separated channel.

| Sequence A | |
|---|---|
| Test Number | Description |
| 1 | Only the optical fiber in channel 1 is connected, unplugging others |
| 2 | Only the optical fiber in channel 2 is connected, unplugging others |
| 3 | Only the optical fiber in channel 3 is connected, unplugging others |
| 4 | Only the optical fiber in channel 4 is connected, unplugging others |
| 5 | Only the optical fiber in channel 5 is connected, unplugging others |
| 6 | Only the optical fiber in channel 6 is connected, unplugging others |
| 7 | Only the optical fiber in channel 7 is connected, unplugging others |
| 8 | Only the optical fiber in channel 8 is connected, unplugging others |

**Sequence B** is generated by analyzing a combination of 2 on 2 channels of the optic fibers after the splitter, thus unplugging the others and using all those connected (Table 2b):

Table 2b – Tests with combination 2 on 2 channels.

| Sequence B | |
|---|---|
| Test Number | Description |
| 1 | Optical fibers in channel 1 and channel 2 connected, unplugging others |
| 2 | Optical fibers in channel 3 and channel 4 connected, unplugging others |
| 3 | Optical fibers in channel 5 and channel 6 connected, unplugging others; |
| 4 | Optical fibers in channel 7 and channel 8 connected, unplugging others |
| 5 | Optical fibers in channel 1, 2, 3 and 4 connected, unplugging others |
| 6 | Optical fibers in channel 5, 6, 7 and 8 connected, unplugging others |
| 7 | Optical fibers in channel 1, 2, 3, 4, 5, 6, 7, 8 connected |

**Sequence C** is generated by analyzing a crescent number of connections (Table 2c):

Table 2c – Tests with a crescent number of connections.

| Sequence C | |
|---|---|
| Test Number | Description |
| 1 | Only the optical fiber in channel 1 and unplugging all others |
| 2 | Optical fibers in channel 1 and channel 2 connected, unplugging others |
| 3 | Optical fibers in channel 1, 2, 3 connected, unplugging others |
| 4 | Optical fibers in channel 1, 2, 3, 4 connected, unplugging others |
| 5 | Optical fibers in channel 1, 2, 3, 4, 5 connected, unplugging others |
| 6 | Optical fibers in channel 1, 2,3,4,5, 6 connected, unplugging others |
| 7 | Optical fibers in channel 1, 2, 3, 4, 5, 6, 7, 8 connected |

### 2.4-Splitter Equation Inference

In this section, we introduce basic knowledge concepts to explain the splitter's equation, used in this research. Analyzing the physical concept of the laser beam emitted by the OTDR equipment, the energy with value $\epsilon$ should be proportional to its respective characteristic oscillation frequency, from one hypothetical oscillator. In 1901, Plank expressed this statement by presenting the proportionality constant $\hbar$ [22].

Next, this energy was associated to the energy of the electromagnetic wave, accounting for energy required to create an electromagnetic field ("quantum"). Thus, this demeanor associated to a unit ("particle"), as opposed to an electromagnetic wave, is referred to as a "photon". Nowadays, Planck´s relation describes the energy of each photon, in terms of photon frequency. This is also known as Plank's relation (Equation 1):

(Eq. 1): $\quad \epsilon = \vartheta \cdot \hbar$

Where:
$\epsilon$ = Beam energy;
$\vartheta$ = Beam frequency (or $1/\vartheta = \lambda$, the wavelength);
$\hbar$ = Plank's constant.

The OTDR laser is a monochromatic beam with $\lambda$ = 1550nm. Consequently, the frequency $\vartheta$ is constant. Thereafter, the energy $\epsilon$ is conserved in the OTDR PON Splitter. Energy density is the amount of energy stored in a given space system or region per unit volume or mass. Anything that can transmit energy can have an intensity associated to it. Therefore, Law of Conservation is focused on intensity, allowing the superimposition of intensities presented in Figure 3:

Next, the verification of the following equation occurs:

(Eq.2):
$$I_0 = \sum_{i=1}^{n} I_i$$
$I_0$ = signal intensity of the incident laser light in the splitter.
$I_i$ = signal intensity in the *i-th* splitter branch.

Equation 2 considers the total output power from all splitter branches, expected to be equal to the input power. It is very important to clarify that there are various types of losses, which affect the performance of the splitter design. For example, losses caused by scattering and absorption at the fiber cladding. In this research, the losses associated with the PON splitter are relatively small and negligible.

Based upon adaptation of the OPLL (Optical Phase Looked Loop) [34] and Interferometry [35], it is also important to consider that the OTDR photo detector acts as a mixer, since the photocurrent is proportional to the intensity of the incident optical signal. Two optical incident fields, occurring on the detector, result in a current that includes a term proportional to the product of the two fields [33]. Figure 4, presents a feedback system that enables electronic control of the output phase of the semiconductor lasers output, which are the basic building blocks of most modern optical communication networks.

Satyan [33] states that the semiconductor laser optical phase-locked loop beam has an output $a_s \cos(\omega_s t + \varphi_s(t))$, and the master laser output is given by $a_m \cos(\omega_m t + \varphi_m(t))$. Equation 3 is proposed by [33] to the detected photocurrent, i.e. an evaluation of two beams targeting the Photodetector (PD):

(Eq. 3)
$$I_{PD}(t) = \rho \cdot \left(a_m^2 + a_s^2 + 2a_s a_m \cos[(\omega_m - \omega_s)t + (\varphi_m(t) - \varphi_s(t))]\right)$$

Here, $\rho$ is the responsiveness of the Photodetector (PD). Taking into consideration the OTDR laser beam, the variation of phase is null. Thus, $2a_s a_m \cos[(\omega_m - \omega_s)t + (\varphi_m(t) - \varphi_s(t))]$ is negligible, due to the behavior of the splitter on the passive optical network. Therefore, the only remaining part is the sum of squares.

When considering OTDR measurements, the signal intensity $P$ with respect to fiber distance $z$, in dB, is a well-known equation $5 \log_{10}(P(z))$ for measures of gain or attenuation given by the ratio of input and output powers of a system, or even by individual factors that contribute to such ratios. In addition, OTDR always shows two-way loss, which means that:
$P_{dB}^{2way} = 5 \log_{10}(P(z)) = \frac{10}{2} \log_{10}(P(z)) = 10 \log_{10}\left(\sqrt{P(z)}\right)$.

Considering the superimposition of powers from different branches $P_c(z)$, the conversion of the power to linear scale is necessary since the addition of power in dB scale is not permitted:
$P_{lin}^{2way} = 10^{\frac{1}{5}P_{c\_dB}} = 10^{\frac{2}{10}P_{c\_dB}} = 10^{\left(\frac{1}{10}P_{c\_dB}\right)^2} = \left(10^{\frac{1}{10}P_{c\_dB}}\right)^2$.

Grounded in these considerations, the total power over distance for all channels (or branches / drop fibers) is the sum of power over the distance of all individual branches, as shown in Equation (4):

(Eq. 4): $$S(z) = 10 \log_{10}\left(\sqrt{\sum_{c=1}^{N} (10^{0.1 P_c(z)})^2}\right)$$

Where:

$P_c$ = OTDR is the measured signal (discrete) per channel $c$ of the Splitter in dB.
$N$ = number of channels of the Splitter.
$z$ is the fiber distance independent variable.

The authors ran validation simulations for this equation within the sequence of tests, as presented in Section 3.3.

The validation process considered the isolated signals measured for each channel in the Sequence A of tests. From these data, the authors simulated several combinations of channels connecting sequences, shown in B and C through the application of Equation 4. Each simulation was then compared to the measured combination thereof. Using Matlab™, the OTDR CSV files were read and respective graphics were generated with the objective of testing the equation and having the results validated. All tests were successful. For other network configurations and other Split ratios a new set of experiments are necessary.

Figure 5, presents one example of the comparisons, where Channel 1 (blue) is measured by OTDR and Equation 4 is simulated as a red curve. Figure 6 presents an example with four channels (Ch. 5, 6, 7 8) and Figure 7 presents all channels to validate the proposed Equation 4.

In all tests, the signal is zoomed from Splitter until the last channel (dead zone). To guarantee the precision, we take Pearson's correlation [36]. For all figures tested, the value is near to 1(one). Its means (mathematically) that both variables are very similar.

The region of interest is between the splitter and the Dead Zone. Before the Splitter, Equation 4 cannot be used and neither in the Dead Zone. As we can see in Figure 5, Figure 6 and Figure 7, the measured and simulated signals are highly coherent.

### 3- Signal Separation

### 3.1-OTDR Measure Simulation

For simulation purposes, we divided the traditional OTDR waveform of a pulse response reflection into the following parts, represented by Figure 8 and described in Table 3.

Table 3: Simulated Signal in Function of sample index.

| Sample id ($\bar{x}$) | $y(\bar{x}) = ...$ | Number of Samples |
|---|---|---|
| $1 \cdots A$ | $y_0 + m.\bar{x}$ | # Samples before peak |
| $A + 1$ | $(y(A) + v_f).\sqrt{2}/2$ | 11 * |
| $A + 2 \cdots B$ | $y(A) + v_f$ | |
| $B + 1 \cdots C$ | $y(B) - 3.86(\bar{x} - B)$ | 4 ** |
| $C + 1 \cdots D$ | $y(C) - 2.41 \ln(\bar{x} - C) + 1$ | # Samples until end of signal |

(*) Common Reflectance for this particular set of measurements, depending on the distance $d$ between each sample. In this work, we use $d$ = 0.5 meters (OTDR configuration).
(**) Dead zone observation of pre-analysis of the signal

In order to simulate the signal of one channel, we need to implement a sort of iterators, which will write amplitudes over an array of the same size, as the number of samples in the measured signal. Based on the network design project and OTDR settings, we already know the location of the splitter. Therefore, it is fair to assume that the proposed algorithm can consider just the samples after the splitter, just for purposes of simulation. Suppose this measured signal has a length (starting right after the splitter) that could define a close interval {1 … D} (illustrated by Figure 8) that addresses the array. In this case, the measured signal is then represented by the simulated amplitudes array y(x):

➢ Within the interval {1…A}, y(x) is defined by a linear equation that reflects the Rayleigh backscattering effect. In the proposed linear equation, m is the fiber slope or fiber attenuation (typically expressed in dB/km). In this work, we use m = atan (-1/1150) based on the manufacturer's datasheet for the experimented fiber and the ITU-T g.652.B standard [12]. Considering pointing the iterator to the next interval, i.e., to change the simulation behavior, the trigger here is the last sample registered before the end of the fiber. This information is also retrievable from the network design project and OTDR settings.

- In the interval {A+1 ... B}, y(x) is defined by a constant added to the amplitude of the signal saved on the array in the last address from the previous interval. This constant represents Fresnel´s reflective event. Fresnel´s reflection results from the discontinuity on indices of refraction observed when a fiber abruptly ends. For example, a Fresnel´s reflection may be caused by the end of a cleaved fiber, an undermined connector, a mated connector or a mechanical splice. Therefore, vf is the raise of amplitude due to reflection, expressed in dB. In this work, we use vf = 21, based on the manufacturer's datasheet for the experimented fiber. Note that the very first sample of the interval ({A+1}) uses a scale factor of approximately 0.707 in order to smooth the transition for the Fresnel reflective event and respect the fact that real world discontinuities are not abrupt. If the distance between samples are large enough, one could dismiss this scale factor. The trigger for changing the simulation behavior is simply after counting up to 11 samples. This proved to be enough considering the equipment and fibers from experimentation and the OTDR's resolution of 0.5m between samples.

After Fresnel´s reflection is considered, the OTDR measured signal has two distinct behaviors [3, 11]. First, there is a linear decline of amplitudes. After, the signal presents an exponential declination:

- That first behavioral concept is simulated within the interval {B+1 ... C}. The simulation y(x) is here defined by a linear equation, starting up to the amplitude of the signal saved on the array in the last address from the previous interval. The authors obtained angular coefficient for this line (constant of -3.86) by approximation from applying linear regression to several experimental measures. The trigger for changing the simulation behavior is simply after counting up to 4 samples. This proved to be enough considering the equipment and fibers from experimentation and the OTDR's resolution of 0.5 m between samples.

- For interval {C+1 ... D}, the second behavioral concept is implemented. The signal is considered an exponential diminishing tail [3],[11]. We obtained the Logarithmic coefficient (constant of -2.41) by approximation from applying logarithmic regression modeling to several experimental measures. This behavior is extended until the end of the array.

The resulting array constitutes the simulated signal. Note that the dead zone corresponds to the interval {A+1 ... D}. One channel per time is simulated, according to some parameters, and Equation 4 is then applied for adding up the simulated signals. The summation of all simulated channels is the final simulated signal to be consider as a mathematical representation of the real measured signal.

**3.2-Diferential Evolutionary Algorithm**

The Differential Algorithm is an evolutionary algorithm proposed in the second half of the 1990s [13-17] for solving optimization problems in continuous variables. Over the last decade, this algorithm has proven to be a successful technique for many applications, such as neural network training, identification of parameters, and design of devices and systems in electrical and electronic engineering. The success of this method is based on the mechanism of the differential mutation operator. This operator generates new solutions starting from vectors constructed with differential pairs of solution candidates drawn from its own population. The differential distribution of vectors in the space of variables for optimization generates search directions with different sizes, which depend on the position of the points used in the construction of the mutant vectors. These directional domains and step sizes fit the characteristics of the problem, with the adaptation properties of self as the population progresses in the direction of the "best" solution.

A differential mutation employs the difference between pairs of individuals in the population to generate the current perturbation vectors called vector differentials. However, as the algorithm progresses in the search process, the spatial distribution of population changes according to the landscape of the objective

```
t ← 1;
Initialize the population X_t = { x_{t,i}, i = 1, … , N};
While stop criteria do
        For each x_{t,i}, i = 1 to N do
                Random Selection
                        r_1, r_2, r_3 ∈ {1, …. , N}
                Random Selection δ ∈ {1, … , N}
                For j = 1 to n do
                        If U_{[0,1]} ≤ C ∨ j = δ_i , do
                                u_{t,i,j} = x_{t,r1,j} + η ( x_{t,r2,j} − x_{t,r3,j})
                        Else  u_{t,i,j} = x_{t,i,j} ;
                End for
        End for
        For each x_{t,i} , i = 1 to N Do
                If f(u_{t,i}) ≤ f(x_{t,i}) do x_{t+1,i} ← u_{t,i}
                        Else x_{t+1,i} ← x_{t,i}
        End For
        t ← t + 1;
End while
```

function or cost function. In turn, this change alters the orientations and sizes of the differential vectors that can be created from the population.

**Pseudo-code of basic differential evolution algorithm (adapted from [14]).**

For this reason, the distribution of the differential vectors, and therefore, the distribution of directions and step sizes of perturbations, fits the landscape function. This characteristic of self-adaptation mutation differential evolution algorithm provides important qualities from the point of view of optimization, such as strength, versatility and efficiency in several problems. Equation 5 illustrates this procedure:

(Eq. 5): $u_{t,i} = x_{t,r1} + \eta ( x_{t,r2} − x_{t,r3})$

with: $r1, r2, r3 \in \{1, \ldots, N\}$. The vector $u_{t,i}$ is the *i-th* mutant solution and $\eta$ a scale factor applied to the differential vector parameter of the differential evolutionary algorithm. The vector $x_{t,r1}$ applied to this differential mutation is called "base vector".

Using this procedure, we obtain a mutant population $V_t = \{v_{t,i} ; i = 1, \ldots N\}$. The next steps in the algorithm are very simple. Individuals in the current population $X_t$ are recombined with individuals of the mutant population, producing offspring or population of test solutions $U_t$.

In this research, we use the classical version of the differential evolution algorithm employing the discrete recombination with probability $C \in [0, 1]$. It is important that N be the number of channels (previously known since the network design is ready and the network built). Index $\delta \in \{1, \ldots, N\}$ is an index drawn for the random test vector *i*. As sometime equality $j = \delta_i$ will be verified, this condition ensures that at least one of the parameters of the test solution will be inherited from the mutant individual.

Parameter $C$ controls the fraction of values in $u_{t,i}$ that are copied from the mutant vector $v_{t,i}$. As closer to 1 the value of $\delta$ becomes, greater is the chance that the test solution contains many values inherited from the mutant vector. At the limit, when $= 1$, the test vector is equal to the mutant vector [17].

This work uses the chromosome representation presented in Figure 9. Note that there is one element for each channel of the Splitter. The $y0_{cn}$ is the initial value of the Power in dB in channel *n* after the splitter.

Each individual $x_{t,i}$ of the population at generation *t* is a possible solution of the problem and, in the present case; structured in the form of the chromosome described in Figure 9. The DE heuristic search for the proper set of $y0_{cn}$ is:

$$\min_{y_{0_{cn}}}(\vec{M} − \vec{S})$$

Where: $\vec{M}$ is the measured OTDR signal and $\vec{S}$ is the superimposition of each one of the *n* simulated channels.

Note that, when this heuristic converges, one obtains the parameters of each separated channel. The quality pattern matching is now measured by Pearson´s correlation between the measured signal (experiments) and the best-simulated signal (algorithm). We have observed repeating the experiments that correlations above *0.97* are a necessary condition to guarantee a satisfactory result. Unfortunately, the computational effort is very high, spending long periods to provide a good solution.

To deal with the performance problem, we implemented some loops of the algorithm in parallel. This technique creates several processes that can compute concurrently on the same loop. A "parfor-loop" can provide a significantly better performance than its analogous for-loop. In our case, we try 2, 3 and 4 processes. We present a panoramic view with results, on table 4. The differential evolutionary algorithm parameters are presented along with the attained correlation and processing time.

Table 4: Evolutionary Algorithm parameters evaluation.

| Sample | Algorithm | Individuals | Generations | Correlation | Time | Method | C | ETA |
|---|---|---|---|---|---|---|---|---|
| 1 | DE | 200 | 600 | 0.98268 | 702.76 | For | 0.3 | .05 |
| 2 | DE | 100 | 600 | 0.97586 | 364.03 | For | 0.3 | .05 |
| 3 | DE | 100 | 400 | 0.9563 | 225,918 | For | 0.3 | .05 |
| 4 | DE | 100 | 400 | 0.97988 | 139.06 | ParFor 4 | 0.3 | .05 |
| 5 | DE | 200 | 400 | 0.97476 | 254.27 | ParFor 4 | 0.3 | .05 |
| 6 | DE | 200 | 600 | 0.98278 | 381.5 | ParFor 4 | 0.3 | .05 |
| 7 | DE | 200 | 600 | 0.95621 | 6.561.784 | ParFor 2 | 0.3 | .05 |
| 8 | DE | 100 | 500 | 0.96321 | 170.35,05 | ParFor 4 | 0.3 | .05 |
| 9 | DE | 100 | 600 | 0.98104 | 208.85 | Parfor 4 | 0.3 | .05 |
| 10 | DE | 100 | 500 | 0.93996 | 271.18 | For | 0.3 | .05 |
| 11 | DE | 200 | 400 | 0.97921 | 453.24 | For | 0.3 | .05 |
| 12 | DE | 1000 | 300 | 0.94641 | 1125.88 | Parfor 3 | 0.3 | .05 |
| 13 | DE | 500 | 1000 | 0.9567 | 1457.00 | Parfor 4 | 0.3 | .05 |

Note that the lowest time with highest correlation is sample 4, which spent *139 seconds* to obtain a correlation of *0,979*. We used an INTEL® processor CORE i7 (2 GHz), with four physical processors for this evaluation.

When we obtain correlations above *0.97* with lower times, we observe that this method is sufficient to guarantee the appropriate solution that proposes the separation of the superimposed OTDR signal after the splitter on the FTTH PON network.

Figure 10 , illustrates the difference between the measured signal (blue) and simulated signal (red) by use of the ED algorithm when the correlation is below *0.97*. In this example, it is possible to identify some peaks of simulated signal that do not fit exactly in the peak of the measured signal.

Note on the left, that the simulated signal is still not able to fit the measured signal and on the right two peaks are close together, requiring more iterations from ED algorithm to guarantee the separation and the distinguished signal.

Figure 11, presents the results obtained when the correlations are above 0.97, with a very close match.

Figure 12, presents from left to right, all OTDR separated channel signals in blue. The after splitter superimposed simulated channels in green and the measured signal in red.

## 4-Contraints Implementation

### 4.1-The Constraint Data Model for FTTH PON Network

In this work, we used Geometric and Engineering Constraints to specify an FTTH Network Design [31]. Geometric Constraint are the existing relationships between two different geometric entities. Engineering Constraints are equations that represent the concepts of physics in electronic components (e.g., Optical Power, Optical Cable losses, Optical Connectors, Splitters, interferences, noise frequencies filters, etc.).

Therefore, one must consider the possibility of modifying some parameters and their propagation, either upstream or downstream over an electronic cable network. This automation is managed by algebraic manipulations (solution of constraint sets) required by the user. Refer to reference [31] for further details.

The hybrid constraints concept is the ideal solution as a method for solving the systems` equations and geometric constraints. These systems rely on numeric and interactive solving techniques [32]. Graph-based models can decompose the equations into smaller sets in order to achieve a more efficient equation work solution.

This approach also allows for direct manipulation of under-constrained FTTH PON network parameters. Using a constraint engine, an evolving solution of the constraint set is locally identified and solved, even with the presence of constraint cycles (simultaneous equations). In this work, we create a database with a characterization of constraint-based approaches for the

OTDR signal and the physical network distribution by coupling both onto a single platform.

Figure 13 shows a constraint-based set that represents the Optical Cable and one Splitter. In the FTTH Scheme Figure13(a) is the physical representation for the connection between one Optical Cable (Feeder) connected to a splitter. Figure 13(b), the Geometric Graph, presents the current relation of geometric constraints for each one, the optical feeder cable C1 and for the splitter S1. Finally, the Equation Graph Figure 13(c) is the particular equation graph for the optical cable E1.

Constraint satisfaction occurs during the network installation phase (start-up), when the network operator carries out a set of OTDR measurements and performs the coupling to the network design.

Geometric and equation constraint satisfaction occurs in the same process as for all network components such as Feeder Cable, Splitter, Branch Cables and others. Although the equations are different for each component, constraint satisfaction is realized after the geometric constraints are satisfied. After defining the splitter position C1, its equation set E1 is placed. The OTDR measurement trace is, in fact, a data entry of the equation graph.

In Fact the system works as a tool to characterize the network. Besides, constraint satisfaction is performed to allow the correspondence of the physical network with the optical one.

### 4.2- Coupling of the OTDR Signal and CAD Network Design Signal

Each separated curve of the OTDR signal is coupled with a CAD environment, bringing to the map (physical representation) the correspondence with the optical signal, in geographic coordinates. A graphic interface was developed (Figure 14), using AutoCAD MAP 2015. Its purpose is to provide an integrated solution between the optical signal and the physical network design installation. Thus, the system supports the complete FTTH graphic design documentation. Physical cable dimensions and geographic coordinates of network equipments are represented in real scale in a FTTH design, located in the city of Uberlandia, State of Minas Gerais, Brazil.

All symbols were created in a library to represent the Telecom market. In addition, its connection rules are similar to real requirements in the field.

Figure 15 presents the FTTH design tool and the OTDR signal viewer. Note that it is possible to have in the same space: the map (physical network) and the separated signals generated by the proposed algorithm.

With this tool, it is possible through the "cursor" to follow the network cable alignment observing in real time, in a specific window, the optical signal coupled, with the physical cable.

If exists multiple optical branches with lengths that are almost similar which impose a dependency on the pulse width and the OTDR resolution, the use of CAD Network Design coupled with the each signal allows to avoid the possibility to mislead any confusion with similar network branches.

### 5- Time Response

The proposed system is composed of several phases described above which are executed in serial sequence. Figure 16 shows how much time at the ONU side is necessary to the following sequence of processing phases. From the left to right, the first phase is the OTDR measurements process followed by the DE Signal Separation, and sequentially followed by the coupling signal with the Network Design in CAD/GIS environment. Sequentially, there is a fail detection phase where the real time signal is compared with the previous OTDR Signal measurement saved in a Relational Database used to compare the difference of measured signals and detect the possible differences in order to give alerts or alarms.

### 6-Treatment of failures

The proposed system provide a signal separation and its coupling with the CAD Network Design. With the signal comparison between several measurements it is possible to detect differences in the signal to support alerts or alarms. However, the system is limited to detect major failures as for example a cutting of some optical cable or some very sharp bend that generates relevant modifications in the detected signal. Other problems such as bending, fiber cracks and performance are not addressed in this work. Any cracks or disconnects will also cause reflections rather than backscattering which is the expected behavior of a normal fiber. To prevent this problem the calibration tool is suggested to avoid its possibility.

### 7-Practical Implementation / Calibration tool

The method proposed in this work, assumes that the length of each branch in the network is known in advance to the central monitoring system, in addition to its GIS position. The authors assume that this data can be obtained during the installation of the fiber at the initial activation tests recommended by quality assurance procedures recommended by regulatory agency tests. A computer based surveillance tool to support the system calibration (figure 17). Through this tool it is possible to compare the field measurements with the design values. Allowing the calibrating of the system. Both in the initial phase of installation of the network also to be carried out periodically during normal network operation. Regarding the fiber aging process that affect the performance of our approach and thus verify the actual lengths found in OTDR measurements of the network and correct it values if it is necessary.

Using this tool, it is possible to compare the lengths of the branches of idealized network in the projects and compare one by one with the OTDR measures in the field. In Figure yy it is possible to observe the system interface developed, the left side, User can open the CAD screen and select branches of the designed network, just below it is possible to select the file measures in the field carried out in real time by the OTDR. On the right side, it is possible to compare each of the channels and adjust the value to the be adopted by the system. Once done this process, the User saves the data to a database containing the calibrated values.

### 8-Results Comparison

To emphasize the features of this development we performed a comparison study considering the system proposed in this research, referred to as NETMAP along with other similar systems such as SANTAD [18] and FPM [32]. Table 5 presents the results.

Table 5- Comparative study.

| Feature | *NET*MAP | SANTAD | FPM |
|---|---|---|---|
| Centralized monitoring from CO | X | X | X |
| Locate network over a CAD / GIS system | X | - | - |
| Separate Superimposition of signal | X | - | - |
| Fault alarm detection | - | X | X |
| Signal simulation features | X | - | - |
| Graphical user interface (GUI) processing capabilities | - | X | - |

This work is still in its early stages and we still have not implemented the routines for Fault detections yet nor implemented a special Graphical interface for use with the operators in the central office. However, the demonstrated capability to separate the superimposed signal and couple it with a graphics network design contributes towards configurations.

### 9-Conclusions

We have presented a novel strategy to simulate the OTDR measurement signal after the PON Splitters. A superimposition equation is proposed in such way that, when joined with a differential evolutionary algorithm, it allows for the determination of the power parcel diverted after the splitters for each channel. It also allows for the separation of the OTRD signal for each channel.

Still, the separated signal of each channel was coupled with a real scaled CAD/GIS network design resulting in a tool to create FTTH PON projects and to contribute to surveillance proceedings.

The authors propose to realize these measurements during the network installation phase.

However, in order to identify new reflective events (excessive bend or cable cut) additional research is required.

The system provides a map with a network design tool that allows the user to evaluate the network coupled with the optical signal.

It is also important to highlight the proposed PON Splitter equation 4, allowing for a computational simulation.

**Acknowledgements**

The authors would like to thank CAPES and FIBERWORK for sponsoring this research work.

References:

[1] Lima, Gerson F. M; Lamounier, Edgard;Barcelos,Sergio "A TEO-Based Algorithm to detect events over OTDR Measurements in FTTH PON Networks", 4TH IEEE LATIN-AMERICAN CONFERENCE ON COMMUNICATIONS 2012 ,7-9 NOVEMBER, CUENCA - ECUADOR.

[2] S. Chabot and M. Leblanc, "Fiber-Optic Testing Challenges in Point-to-Multipoint PON Testing ", EXFO Application Note 110, http://documents.exfo.com/appnotes/anote110-ang.pdf

[3] Dennis Derickson; "Fiber optic test and measurement"; Prentice Hall PTR, 1998-Michigan University; ISBN9780135343302.

[4] N. Ferrari, L. Greborio, F. Montalti, P. Regio, G. Vespasiano, "OTDR Characteristics for PON Measurements", Proceedings of the 57th International Wire & Cable Symposium, November 9-12, 2008, Rhode Island, USA, pag. 27-35.

[5] Mario Simard, "OTDR PON Ton Testing: the Challenges - The Solution", EXFO Application Note 201, http://documents.exfo.com/appnotes/anote201-ang.pdf

[6]FTTH Council, "FTTH Council - Definition of Terms", available: ttp://www.ftthcouncil.org/UserFiles/File/FTTH_definitions.pdf, 2006.

[7] D.R. Anderson, L. Johnson, and F.G. Bell, "Chapter 10: Analyzing Passive Networks Containing Splitters and Couplers", Troubleshooting Optical Fiber Networks Understanding and Using Optical Time-Domain Reflectometers, California, US: Academic Press, Elsevier, pp. 279-290, 2004.

[8] Yeh, Chien Hung, et al. "Reliable tree-type passive optical networks with self-restorable apparatus." Optics Express 16.7 (2008): 4494-4498.

[9] Bakar, A. A. A., et al. "A new technique of real-time monitoring of fiber optic cable networks transmission." Optics and Lasers in Engineering 45.1 (2007): 126-130.

[10] Triple-Play Service Deployment "A Comprehensive Guide to Test, Measurement, and ServiceAssurance", "JDSU", http://www.ccm.ch/files/jdsu_tripleplaybook_oct200700.pdf

[11] Duwayne R. Anderson, Larry Johnson, Florian G. Bell,"Troubleshooting Optical-Fiber Networks" –Printed book –Elsevier - ISBN: 0-12-0586614

[12] Recommendation G.652 – (11/2009), Characteristics of a single-mode optical fiber and cable: http://www.itu.int/rec/T-REC-G.652-200911-I/en

[13] Chakraborty, U. K. (ed.) (2008). Advances in Differential Evolution, Studies in Computational Intelligence, Springer.

[14] Price, K. V., Storn, R. M. and Lampinen, J. A. (2005).Differential Evolution: A Practical Approach to Global Optimization, Natural Computing Series, Springer.

[15] Rocca, P.; Oliveri, G.; Massa, A. (2011). "Differential Evolution as Applied to Electromagnetics". IEEE Antennas and Propagation Magazine 53 (1): 38–49. doi:10.1109/MAP.2011.5773566.

[16] Storn, Rainer, and Kenneth Price. "Differential evolution–a simple and efficient heuristic for global optimization over continuous spaces." *Journal of global optimization* 11.4 (1997): 341-359.

[17] Kazovsky, Leonid G., "Phase- and polarization-diversity coherent optical techniques," Lightwave Technology, Journal of , vol.7, no.2, pp.279,292, Feb 1989 doi: 10.1109/50.17768

[18] MS Ab-Rahman, B Ng, K Jumari; " Remotely Control, Centralized Monitoring and Failure Analyzing in PON"; IJCSNS International Journal of Computer Science and Network Security, VOL.9 No.2, February 2009

[19]Boonchuan Ng, Mohammad Syuhaimi Ab-Rahman and Kasmiran Jumari (2010). Graphical User Interface for PON Network Management System, User Interfaces, Rita Matrai (Ed.), ISBN: 978-953-307-084-1, InTech.

[20] Hestenes, David. "Toward a modeling theory of physics instruction."*American journal of physics* 55.5 (1987): 440-454.

[21] Roychoudhuri, Chandrasekhar. "The locality of the superposition principle is dictated by detection processes." Physics Essays 19.3 (2006): 333-354.

[22] Holbrow, C. H., E. Galvez, and M. E. Parks. "Photon quantum mechanics and beam splitters." American Journal of Physics 70 (2002): 260.

[23] Recommendation G.650.1, Definitions and test methods for linear, deterministic attributes of single-mode fibre and cable. http://www.itu.int/rec/T-REC-G.650.1-201007-I/en

[24] Pirich, Ronald, and John Mazurowski. "Engineering of fiber optics infrastructure." *Systems, Applications and Technology Conference (LISAT), 2012 IEEE Long Island*. IEEE, 2012.

[25] Bahrampour, Ali Reza, and Fatemeh Maasoumi. "Resolution enhancement in long pulse OTDR for application in structural health monitoring." *Optical Fiber Technology* 16.4 (2010): 240-249.

[26] N. Gagnon, A. Girard, and M. Leblanc, "Considerations and Recommendations for In-Service Out-of-Band Testing on Live FTTH Networks," in Optical Fiber Communication Conference and Exposition and The National Fiber Optic Engineers Conference, Technical Digest (CD) (Optical Society of America, 2006), paper NWA3.

[27] Prasad, Sudhakar, Marlan O. Scully, and Werner Martienssen. "A quantum description of the beam splitter." Optics communications 62.3 (1987): 139-145.

[28] Roychoudhuri, Chandrasekhar. "Reality of superposition principle and autocorrelation function for short pulses." *Proc. SPIE*. Vol. 6108. 2006.

[29] The Fundamentals of an OTDR - EXFO
Available at: "http://www.exfo.com/glossary/optical-time-domain-reflectometer-otdr

[30] LIMA,G.F.M.;LAMOUNIER,JR;CARDOSO.A;"Constraint-based Techniques to Support Electronic TV Network Design" Lambert Academic Publisher, 2010.108p.

[31] LAMOUNIER JR, E.; "An Incremental Constraint-based Approach to Support Engineering Design", PhD Thesis, School of Computer Studies, Leeds University, UK, 1996.

[32] Urban, P.J.; Vall-llosera, G.; Medeiros, E.; Dahlfort, S., "Fiber plant manager: an OTDR- and OTM-based PON monitoring system," *Communications Magazine, IEEE*, vol.51, no.2, pp.S9,S15, February 2013 doi: 10.1109/MCOM.2013.646118

[33] Satyan, Naresh (2011) Optoelectronic control of the phase and frequency of semiconductor lasers. Dissertation (Ph.D.), California Institute of Technology. http://resolver.caltech.edu/CaltechTHESIS:04292011-221312708

[34] LANGLEY, L. et al. Packaged semiconductor laser optical phase-locked loop (opll) for photonic generation, processing and transmission of microwave signals. Microwave Theory and Techniques, IEEE Transactions on, IEEE, v. 47, n. 7, p. 1257–1264, 1999.

[35] DEMTRODER, W.; TITTEL, F. K. Laser spectroscopy: basic concepts and instrumentation. Optical Engineering, International Society for Optics and Photonics,v. 35, n. 11, p. 3361–3362, 1996.

[36] Cox, D.R., Hinkley, D.V. (1974) *Theoretical Statistics*, Chapman & Hall (Appendix 3) ISBN 0-412-12420-3

FIGURE CAPTIONS:

Figure 1 – Proposed Methodology.

Figure 2 – Experimental Network Tested (Splitter configuration and branches).

Figure 3 – Law of Conservation focused on intensity.

Figure 4 – A heterodyne semiconductor laser optical phase-locked loop (SATYAN, 2011)[33].

Figure 5 – Measured Signal (channel 1) x Signal Result using the Equation in one channel.

Figure 6 - Measured Signal (channels 5, 6, 7, 8) x Signal Result using the Equation in four channels

Figure 7- Measured Signal all channels x Signal Result using the Equation in all channels.

Figure 8- OTDR signal simulation pulse parts.

Figure 9 - Chromosome representation.

Figure 10- Results with correlation of 0.95126 (Red) x Measured Signal (Blue).

Figure 11- Signal Measured (blue) x Simulated Signal (Red) Correlation above 0.97

Figure 12- Plots of the separated channel (blue), simulated signal (green), measured signal (red).(3D)

Figure13– Constraints and FTTH PON Network.

Figure 14 - FTTH Design tool (GUI) and a FTTH design sample.

Figure 15- FTTH real design coupled with the OTDR signal (separated after the splitter).

Figure 16- Time response.

Figure17-Calibration Tool

Figures

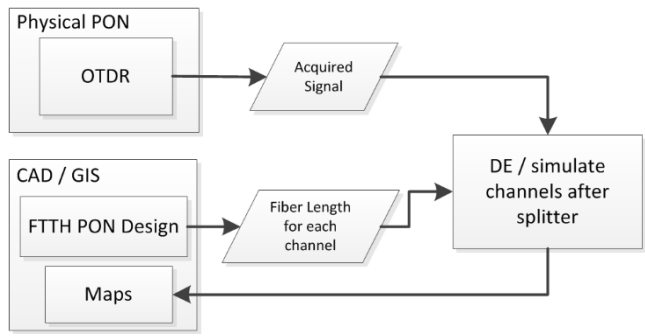

Figure 1

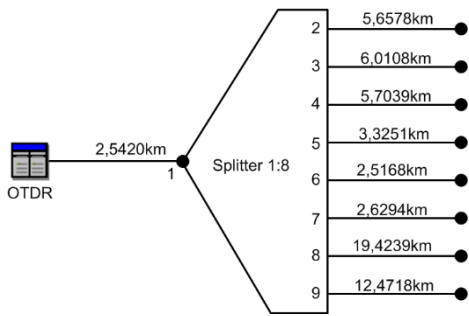

Figure 2

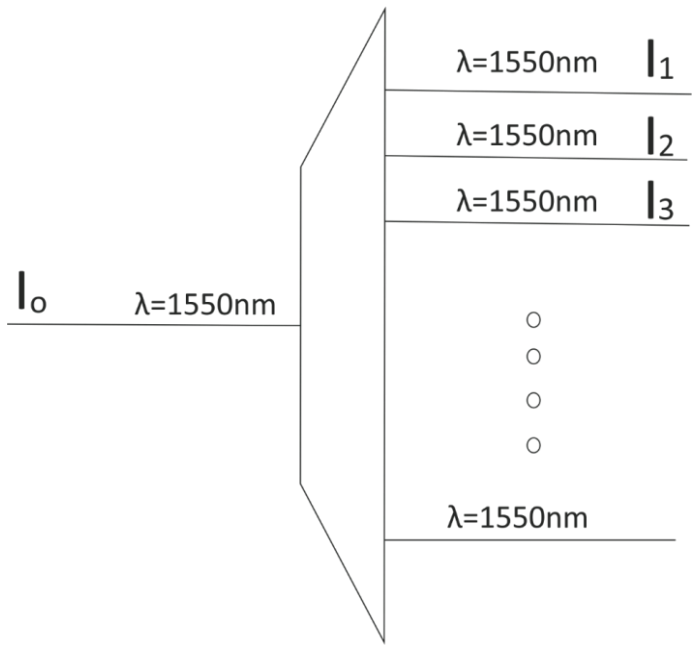

Figure 3

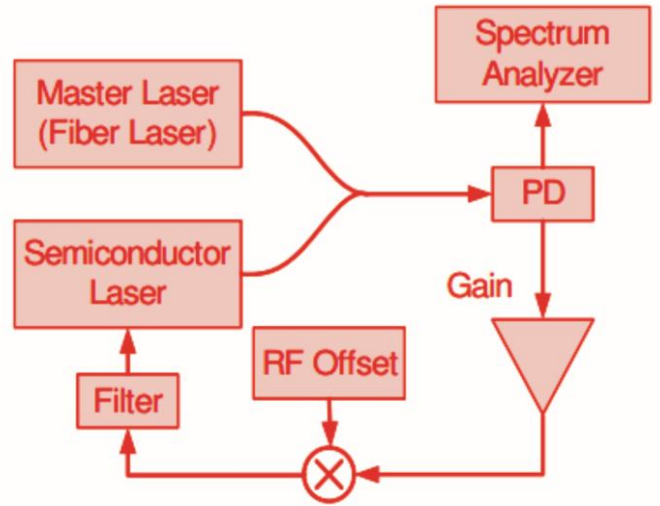

Figure 4

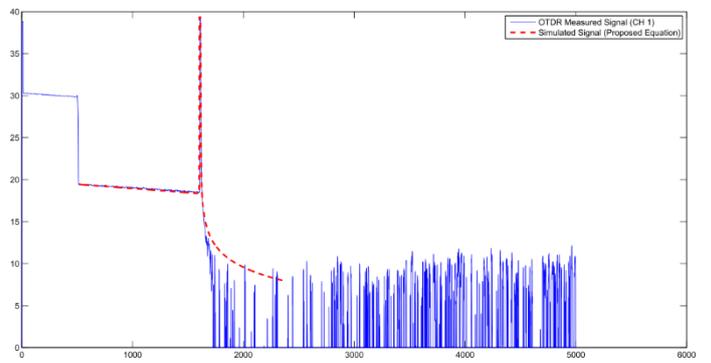

Figure 5

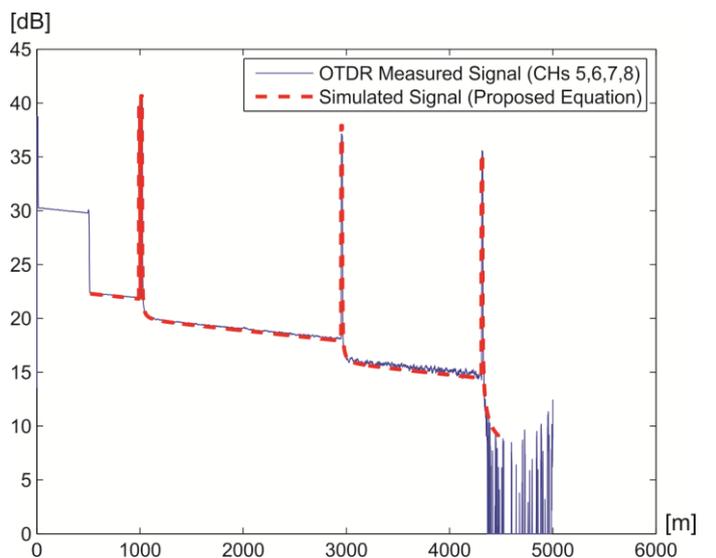

Figure 6

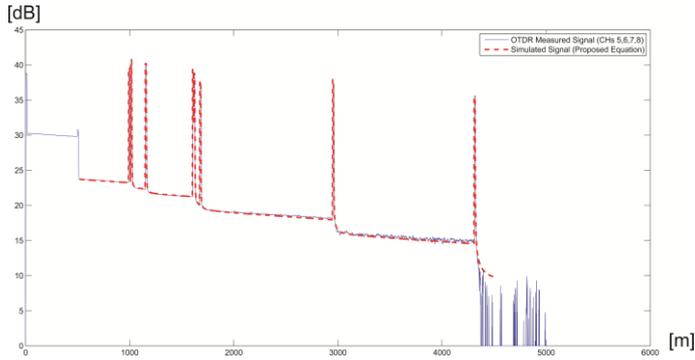

Figure 7

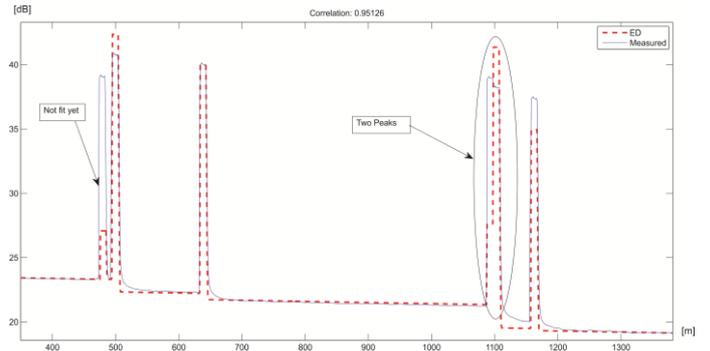

Figure 10

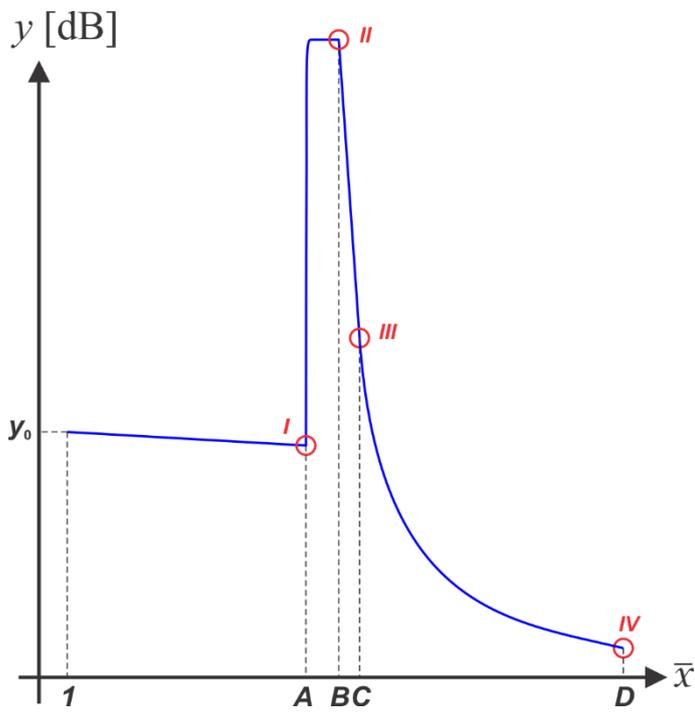

Figure 8

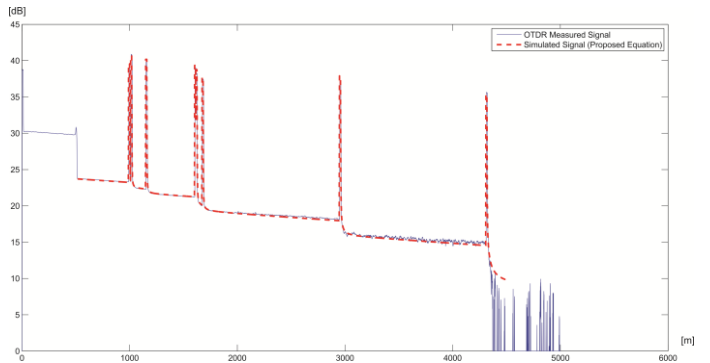

Figure 11

| Channel 1 | Channel 2 | ... | Channel n |
|-----------|-----------|-----|-----------|
| $y_{0_{c1}}$ | $y_{0_{c2}}$ | ... | $y_{0_{cn}}$ |

Figure 9

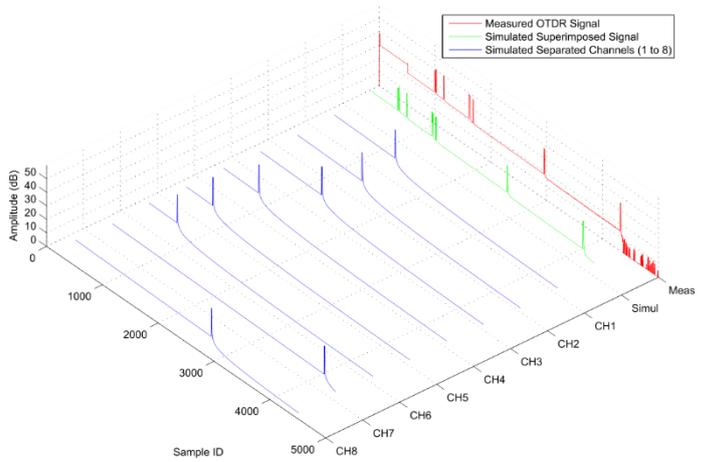

Figure 12

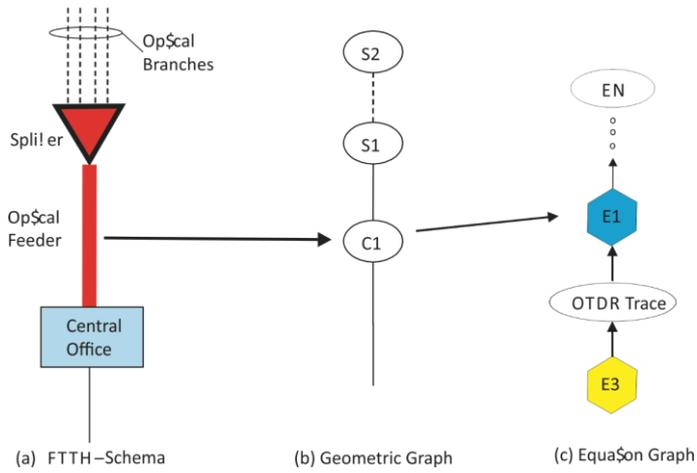

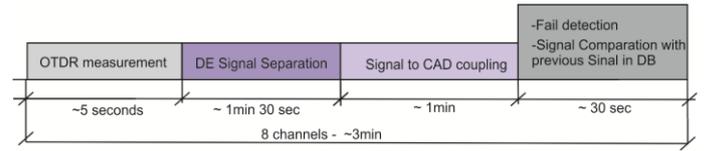

Figure 16

Figure 14

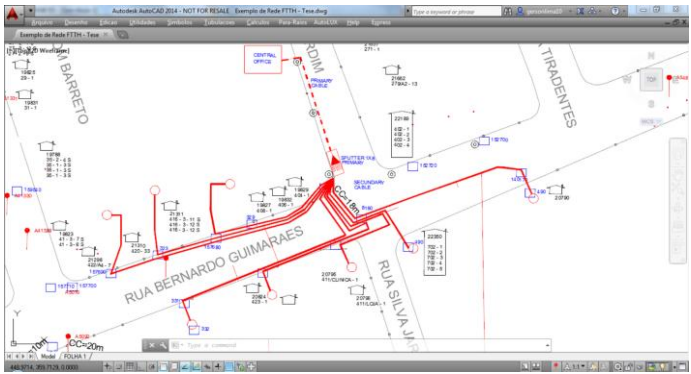

Figure 17

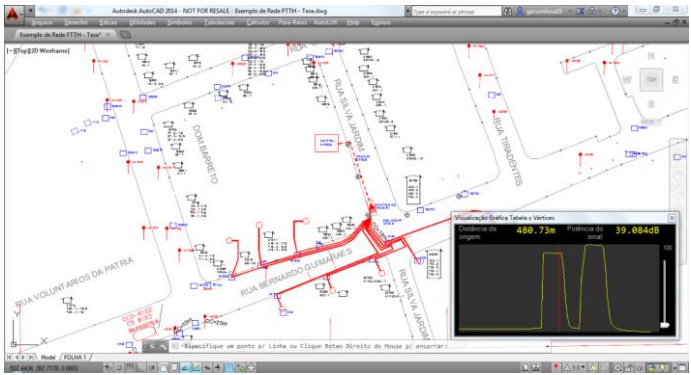

Figure 14

Figure 15